\documentclass[a4paper,showkeys,floatfix,aps,pre,longbibliography,superscriptaddress,reprint]{revtex4-1}
\usepackage{graphics,graphicx}
\usepackage{amsmath,amssymb}

\usepackage{graphics,graphicx}
\usepackage[utf8]{inputenc}
\DeclareUnicodeCharacter{2212}{\ensuremath{-}}
\usepackage{dcolumn,bm}
\usepackage{psfrag}
\usepackage{xstring}
\usepackage{color}
\usepackage{soul}
\newcommand{\srm}
{\affiliation{Department of Physics, SRM University - AP, Amaravati,
 Andhra Pradesh - 522240, India}}

\newcommand{\srmcse}
{\affiliation{Department of Computer Science and Engineering, SRM University - AP, Amaravati,
 Andhra Pradesh - 522240, India}}

 \newcommand{\saha}
 {\affiliation{Condensed Matter Physics, Saha Institute of Nuclear Physics, Kolkata
700064, India}}

\begin{document}

\title{Classical Annealing of Sherrington-Kirkpatrick Spin Glass Using
Suzuki-Kubo Mean-field Ising Dynamics}

\author{Soumyaditya Das}

\email{soumyaditya\_das@srmap.edu.in}
\srm
\author{Soumyajyoti Biswas}

\email{soumyajyoti.b@srmap.edu.in}
\srm
\srmcse
\author{Bikas K. Chakrabarti}
\email{bikask.chakrabarti@saha.ac.in}
\saha

\begin{abstract}
    We propose and demonstrate numerically a fast classical
annealing scheme for the Sherrington-Kirkpatrick (SK) spin glass model,
employing the Suzuki-Kubo meanfield Ising dynamics (supplemented by a
modified Thouless-Anderson-Palmer reaction field). The resultant dynamics,
starting from any arbitrary paramagnetic phase (with local magnetizations $m_i=\pm 1$ for the $i^{th}$
spin, and the global magnetization $m=0$), takes the system  quickly to
an appropriate state with small local values of magnetization ($m_i$)
commensurate with the (frustrated) interactions. As the temperature
decreases with the annealing, the configuration practically remains (in an effective
adiabatic way) close to a low energy configuration as the magnitudes of $m_i$'s and the spin
glass order parameter $q$ grow to unity. While the configuration reached by
the procedure is not the ground state, for an $N$-spin SK model (with $N$
up to 10000) the deviation in the energy per spin $E^0_N - E^0$ found by the
annealing procedure scales as $N^{-2/3}$, with $E^0  = -0.7629\pm 0.0002$,
suggesting that in the thermodynamic limit the energy per spin of the
low energy configurations converges to the ground state of the SK model
(analytical estimate being $E^0 =-0.7631667265 \dots$),
fluctuation $\sigma_N $ in $E^0_N$ decreases as $\sim N^{-3/4}$ and the annealing time $\tau_N \sim N$, making this protocol highly efficient in estimating the ground state of the SK model.

\end{abstract}

\maketitle

\section{Introduction}
The Sherrington-Kirkpatrick (SK) \cite{sher,panchenko} model was proposed as the infinite dimensional or mean field limit of Ising spin glasses having
disordered competing interactions with irreducible frustrations. The model
proved to have unusually complex structure, dynamics and a vast field of applications from materials science to various optimization problems (see e.g., \cite{binder} for
a review). The complex nature of the SK model ground state structure was
identified by Parisi \cite{parisi1,parisi2} to have Replica Symmetry Breaking (RSB). It was
shown that the complete RSB estimated ground state energy per spin $E^0$
of the SK model is $ - 0.7631667265 \dots$ \cite{oppermann,stefan}. Since then, major
numerical efforts have been made (see e.g., \cite{kim,boe_epjb2005,stefan2,kirk,rakch,erbal}), using various
(conformational space) annealing, quenching, and local search techniques that found the difference from the ground state energy per spin $E^0_{N} - E^0$ disappears following the finite size scaling behavior $N^{-2/3}$ for $N$-spin SK glass (see e.g. \cite{boe_epjb2005}) and the
fluctuations $\sigma_N$ in $E^0_{N}$ decreases as $N^{-3/4}$ (see e.g. \cite{bouchaud,kobe}). 

Finding the ground state energy for the SK spin glass for a given configuration is NP hard, and therefore, the approximate algorithms are either parameter sensitive (e.g., the genetic algorithm \cite{proc}) or less accurate (e.g., simulated annealing \cite{grest}) or time consuming (e.g., Extremal Optimization that scales as $N^4$ \cite{boe_epjb2005}). However, given the wide ranging applications and importance of the model \cite{mezard_book}, searching for a computationally beneficial algorithm in finding the ground state of the SK model has remained an outstanding and widely investigated problem through decades (see e.g., \cite{mezard}). 

In this work, we report a classical annealing approach for the SK model using the Suzuki-Kubo mean-field dynamics \cite{suzuki}. As a consequence of the dynamical equations, the individual spins are discrete ($\pm 1$) in the beginning and at the end of the dynamics, but during the annealing process they become continuous variables in $(-1,1)$. As a result, the corrugated free-energy landscape is smoothened during the dynamics and the spins find the appropriate low energy state configuration using a considerably simpler algorithm (cost $N^3$).  However, in Ref. \cite{montanari} an algorithm of cost $N^2$ was proposed. In spite of the simplifications of the dynamics here, the above mentioned finite size scalings are recovered in the ground state. We first describe the Suzuki-Kubo formalism in the context of the SK spin glass. Then we show that the equilibrium properties of the model (phase diagram) can be recovered using this. Then we go on to show how classical annealing can help in finding the ground state of the model using a very low energy state estimate for a finite system of size $N$ in a time that scales as $N^3$. 

\section{Suzuki-Kubo dynamics in Sherrington-Kirpatrik spin glasses}
In this section, we discuss the implementation of the Suzuki-Kubo dynamics for the Sherrington-Kirpatrik spin glass. We give the general formalism for a time dependent temperature. However, we then subsequently discuss the constant temperature case (equilibrium properties) and then the time dependent temperature case (annealing). 

\begin{figure}[tbh]
\includegraphics[width=8cm]{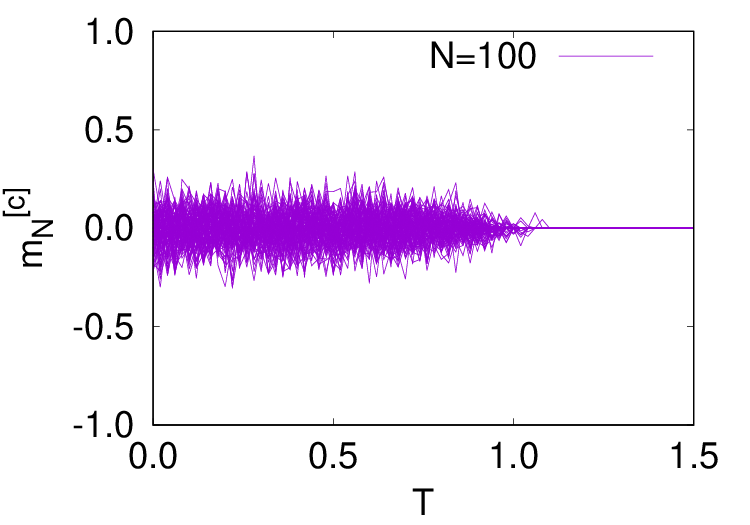}
\includegraphics[width=8cm]{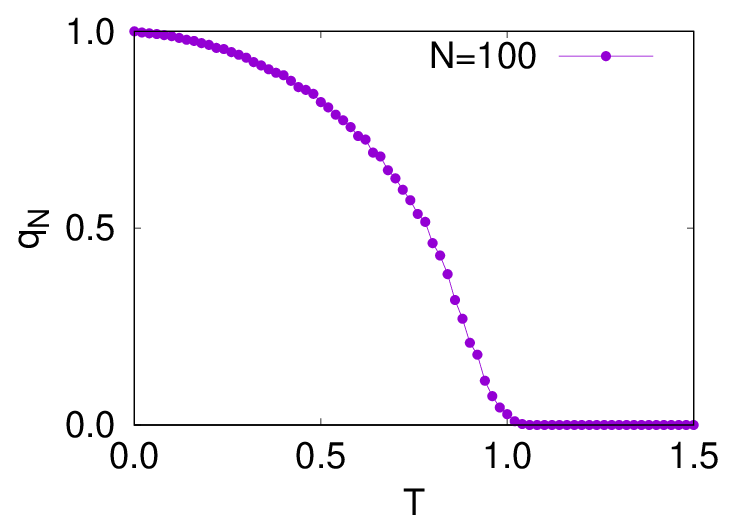}
\caption{The top figure shows the magnetization for different configuration as a function of temperature. It is scattered around zero below the transition temperature $T_g=1$ and becomes identically zero above it. The figure at the bottom shows the spin glass order parameter, which starts with $q=1$ at $T=0$ and becomes zero at $T=T_g$.}
\label{en0}
\end{figure}
\begin{figure*}
\includegraphics[width=16cm]{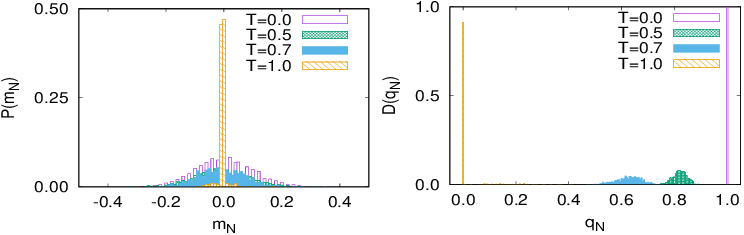}
\caption{The distributions $P(m_N)$  and $D(q_N)$ of the average magnetization $m_N$ (left) and the spin glass order parameter $q_N$ (right) for different values of constant temperatures. The system size is $N=100$. At $T=0$, the local magnetization values could only be $\pm 1$, with almost equal probability, hence the distribution is a Gaussian peaked at zero. For intermediate temperatures, a somewhat broader peak could be seen around zero. But for $T=T_g=1.0$, the distribution is almost a delta function at zero, indicating all local magnetization values are identically zero just above this temperature (see Fig. \ref{en0}), as a consequence of the Suzuki-Kubo dynamics. The associated features are also seen for the distributions of the spin glass order parameter.}
\label{bb}
\end{figure*}

\begin{figure*}[tbh]
\includegraphics[width=17cm]{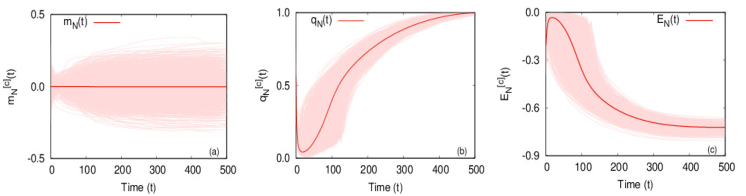}
\caption{The time variations (due to annealing) of (a) magnetization $m_{N}^{[c]}(t)$, (b) spin glass order parameter $q_{N}^{[c]}(t)$ and (c) energy $E_{N}^{[c]}(t)$ are shown. In all these cases the configurational averages are shown in red (thick) line and each of the 1000 configurations are shown in light background. While the average magnetization remains close to zero throughout the dynamics, the magnitudes of the local magnetization are initially lowered, which is reflected in the initial dip of the spin glass order parameter. At later times, the magnitudes of the local magnetization and consequently that of the spin glass order parameter reach unity. The system size here is $N=100$.}
\label{combine}
\end{figure*}

\begin{figure}[tbh]
\includegraphics[width=9cm]{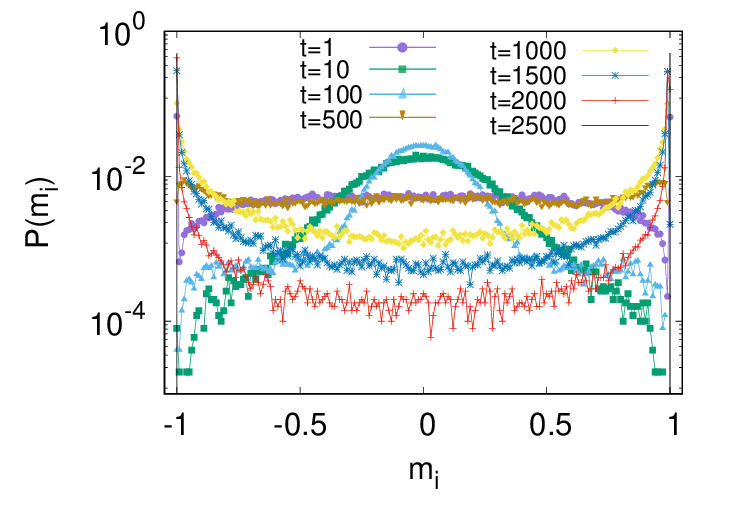}
\caption{The time evolution of the distribution of the individual magnetizations $m_i$ during the annealing dynamics are shown. At earlier times, the distribution is peaked around $\pm 1$, then it shows peak at 0 in the intermediate times, before finally peaking again near $\pm 1$. Exactly at $t=0$ and at $t=\tau$ the distributions are delta functions at $\pm 1$ (since $m_i$ for all $i$ are $\pm 1$ in both cases).}
\label{mz_t}
\end{figure}

\begin{figure}
\includegraphics[width=8cm]{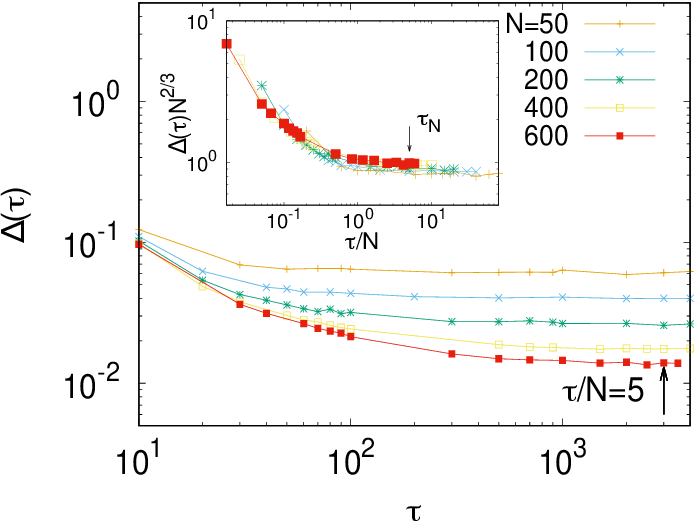}
\caption{The variation of $\Delta(\tau) = E^0_{N}(\tau)-E^0$ is shown with different annealing time ($\tau$) for different system sizes. Clearly $\Delta$ becomes independent of $\tau$ (and $E^0_N(\tau)$ become $E^0_N$) when $\tau > \tau_N$, where the annealing time $\tau_N$ is linear in $N$. }
\label{tau_comb}
\end{figure}
The Hamiltonian \cite{kirk} of the SK model reads
\begin{equation}
 H = - \sum_{<ij>} J_{ij} s_i s_j   
 \label{hamiltonian}
\end{equation}
 where $<ij>$ denotes the distinct pairs (counted once) of the
Ising spins $s_i$, $i = 1, 2, \dots, N$, interacting with long-range
interactions $J_{ij}$. The $J_{ij}$ values are taken randomly from a Gaussian distribution centered at zero

\begin{equation}
P(J_{ij}) = (1/J) (N/2\pi)^{1/2} \exp [-(N/2)(J_{ij}/J)^2] 
\end{equation}

 where, $[J_{ij}^2]_{av}-[J_{ij}]_{av}^2=J^2/N=1/N$.

 The above Hamiltonian, for the purposes of the Suzuki-Kubo mean field dynamics \cite{suzuki}, can be written  as  $H^{[c]}= \sum_i h_i^{[c]} m_i^{[c]}$, for a particular configuration [c] corresponding to a particular realization of the distribution of $J_{ij}$, where $m^{[c]}_i
=\hspace{0.1cm}<s^{[c]}_i>$ where $<.>$ denotes the thermal average and $h_i^{[c]}$ is the effective field faced by the local magnetization $m_i^{[c]}$ at any site $i$. The effective field should, in general, incorporate the Thouless-Anderson-Palmer (TAP) reaction field \cite{thouless,binder} 
for a particular configuration [c]:
\begin{equation}
 h^{[c]}_i (t) = -\sum_{j} J_{ij} m^{[c]}_j(t) - [1 - q^{[c]}(t)]m^{[c]}_i(t),   
 \label{heff}
\end{equation} 
where the spin glass order parameter $q^{[c]}=1/N\sum_{i=1}^N (m_i^{[c]})^2$.  The reaction term is modified here by normalizing the $J_{ij}^2$ \cite{mezard} and by removing the $1/T$ term to avoid its
divergence in the $T = 0$ limit,  as required for its extension
(see e.g., \cite{ishii,ray}) to the quantum case (SK model in
transverse field $\Gamma$ at $T = 0$). The other part of the
TAP reaction field  ($[1- q^{[c]}]m$) is perfectly general in both classical as well as quantum cases and disappears only
as  the  spin glass ground state approaches  ($q \to 1$), in both the cases.

 In general, the classical annealing dynamics is introduced as well, if we keep the temperature to be time dependent $T\equiv T(t)$.  Then we can lower the temperature from a value at or above the spin glass transition point temperature  $T_0 \ge T_g = 1$ \cite{binder} down to zero  in annealing time $\tau$: $T(t)=T_0\left(1-t/\tau\right)$.
 The dynamical equation for the local magnetizations then reads
\begin{equation}
 dm^{[c]}_i/dt = - m^{[c]}_i  + \tanh (h^{[c]}_i/T(t))
\end{equation}
\noindent in discrete time form
\begin{equation}
 m^{[c]}_i(t +1) = \tanh \left[h^{[c]}_i(t)/T(t)\right].
 \label{mag}
\end{equation}

The above equations can be iterated, starting with $m^{[c]}_i(t=0)=\pm 1$ with equal probability and one time step consisting of $N$ randomly chosen updates ($N$ being the system size). 

\subsection{The case of constant temperature}

Given that the Suzuki-Kubo dynamics has not been applied to the SK model before, it is important to note before we proceed with the results of annealing here that one can use the above equations, for constant $T$ (not a function of time), to recover the equilibrium properties of the SK model.

For simulating equilibrium properties of the model, we start with a random initial condition, where initially all $m_i^{[c]}$ are assigned $\pm 1$ with equal probability. After that within each time step, one site is chosen at random and is updated using
$m^{[c]}_i(t +1) = \tanh \left[h^{[c]}_i(t)/T\right]$,
where, $h^{[c]}_i (t)$ is given by Eq. (\ref{heff}) and the only change in the time evolution equation for $m^{[c]}_i(t)$ here and Eq. (\ref{mag}) is that here the temperature $T$ is fixed. $N$ such updates constitute one time step. We continue until in the successive steps the changes in the magnetization values $m_i^{[c]}(t)$ (for all $i$) fall below a pre-assigned threshold $|m_i^{[c]}(t+1)-m_i^{[c]}(t)|<\delta$, with $\delta=0.0001$. These 
values are then averaged over space (for all values of $i$) to find the  average magnetization $m_N^{[c]}$ for a given configuration $[c]$ of system size $N$ and its average over configurations gives $m_N$. Similarly, the average spin glass order parameter is denoted by $q_N$  (shown in Fig. \ref{en0}). For different configurations, the spatial average of the local magnetization $m_i^{[c]}(t)$, after equilibrium is reached (denoted by $m_i^{[c]}$), is scattered around zero for $T<T_g=1$, and all $m_i^{[c]}(t)$ (and by extension $m_N^{[c]}$) become identically zero for $T>T_g$, as a consequence of the Suzuki-Kubo equation. The spin glass order parameter starts from unity at $T=0$ and vanishes, as expected, at $T=T_g$. However, as local magnetization values are identically zero for $T>T_g$, the spin glass order parameter also is identically zero here (no scatter around zero). 
This behavior is further clarified by the distributions of the values of $m_N$ and $q_N$ ($N$ being the system size) as shown in Fig. \ref{bb} for different temperatures after reaching equilibrium.  

The above analysis shows that the Suzuki-Kubo dynamics can reproduce the equilibrium (at constant $T$) phase diagram of the SK spin glass. The only difference is that the local magnetization magnitudes can now take continuous values (as opposed to discrete $\pm 1$ only). However, that does not affect the transition temperature and, as we shall soon see, the $T\to 0$ limit of the model, where the discreteness is recovered. However at finite temperature since our algorithm does not contain
the factor $1/T$ in the reaction term, we do not expect the energy and
overlap to converge to the correct equilibrium values described by the
TAP equations.


\subsection{Annealing of SK spin glass with Suzuki-Kubo dynamics}
Our interest here, of course, is to examine the ground state ($T\to 0$)  properties of the SK spin glass, as obtained through annealing using Suzuki-Kubo dynamics. Specifically, how might the Suzuki-Kubo dynamics help in reaching a low energy state in a shorter time (see also \cite{cavanga} for the nature of such states from TAP equation). 

As for simulating the annealing dynamics, we have used Eqs. (\ref{mag}) and (\ref{heff}) for $N$
up to 10000 from $t=0$ to $t=\tau$ with time varying temperature, as mentioned before. We report the finite size scaling of the very low energy state estimates as they approach the universal \cite{carmona} ground state energy for the systems, the finite size scaling of the fluctuations in the energy values to follow the standard scaling relations mentioned above. Surprisingly, the algorithmic cost here scales as $N^3$.

In simulating the model with the Suzuki-Kubo dynamics following  Eqs. (\ref{mag}) and (\ref{heff}), the initial temperature is chosen at $T_g=1$ and is lowered linearly such that at time $t=\tau$ it reaches zero. The initial local magnetisation values, for a given configuration are assigned to be $\pm 1$ with equal probability. A local magnetization is then randomly selected and updated following Eqs. (\ref{mag}) and (\ref{heff}) and $N$ such updates constitute one time step, just as before.

As indicated above, we denote the configurational average of total magnetization and the spin glass order parameter as $m_N(t)=\left[m^{[c]}_N(t)\right]_{av}$ and $q_N(t)=\left[q^{[c]}_N(t)\right]_{av}$ respectively, $[.]_{av}$  implies the configurational average (different configurations start with different set of initial spins and $J_{ij}$).

\begin{figure}
\includegraphics[width=8.5cm]{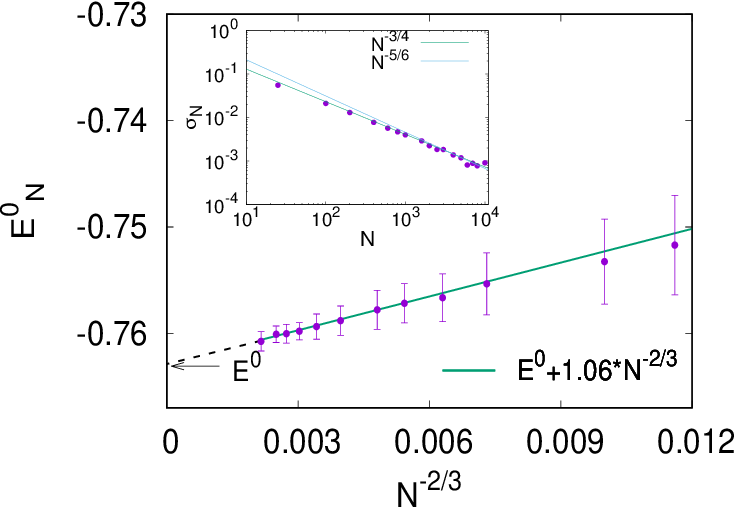}
\caption{The lowest energy values for given system size are plotted against $N^{-2/3}$ which shows a scaling $E_{0}(N)\sim N^{-2/3}$. The ground state energy ($N \rightarrow {\infty}$) is the Parisi value $E^0=-0.7631 \dots$. From the least-square fitting we get a ground state energy which is $E^0=-0.7629\pm 0.0002$ (considering the exponent to be 2/3). The inset shows the variation of the fluctuations $\sigma_N$ of
$E^0_N$. It appears that while for the entire range of $N$ values we considered here, $\sigma_N \sim N^{−3/4}$ gives good fit, for larger values of $N$, $\sigma_N \sim N^{−5/6}$ (cf. Ref. 
\cite{parisi3,parisi4}) gives perhaps a better fit.}
\label{en}
\end{figure}

\begin{figure}
\includegraphics[width=8.5cm]{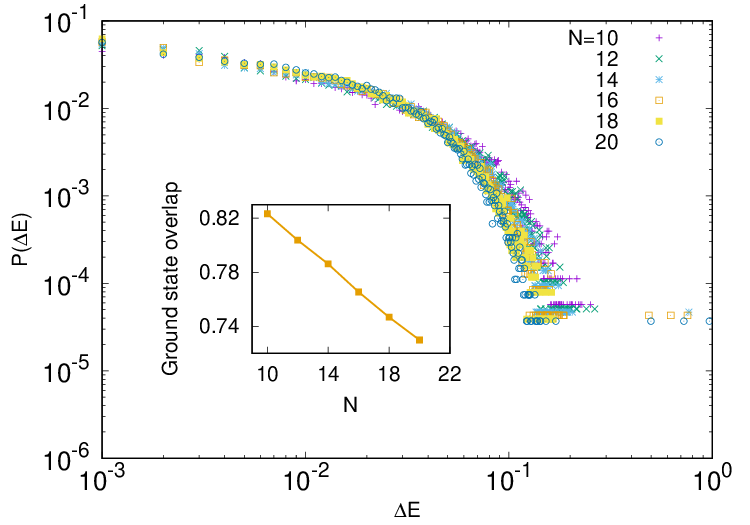}
\caption{Comparisons of ground state energies and that obtained through our algorithm are shown. The main figure shows the
probability distribution of the differences between the energies obtained by this algorithm and that of the actual ground states (obtained through brute force method). The inset depicts the fraction of cases the algorithm finds the true ground states as a function of the system size.}
\label{gnd}
\end{figure}

During the dynamics, the individual local magnetization values $m_i^{[c]}(t)$ are significantly lowered initially. The effect of this can be seen in Fig. \ref{combine}(a). Also the time dependent distribution  of the individual local magnetization values shows that more clearly (see Fig. \ref{mz_t}). This means that the effective cooperative energy barriers separating different configurations of the local magnetizations are temporarily lowered. Therefore, the probability that a (local magnetization) configuration gets trapped in a local minimum of the energy landscape, is reduced. As $t$ increases, the system finds the lower energy states and the orientations of $m_i^{[c]}(t)$ gets adjusted accordingly, still maintaining, on average, a much smaller magnitude than unity. Subsequently, the local magnetization values increase gradually to $\pm 1$ (see Fig. \ref{combine}), again restoring the large energy barriers and prohibiting further changes in the local magnetization orientations, hence the spin glass order. However, the local magnetization orientations are frozen at a configuration that is already adjusted to be a very low energy state. 

\begin{table}
\caption{The number of configuration averages done for each
system size ($N$) and the corresponding estimates for the ground state energies ($E^0_N$)  and their errors (estimated from
the standard deviations) are listed below.}
    \centering
    \setlength{\tabcolsep}{5.0pt} 
\renewcommand{\arraystretch}{1.0}
    \begin{tabular}{|l|c|r|}
    \hline
        System size ($N$) & Configs. & Final energy ($E^0_{N}$) \\ 
        \hline
$25$ &$10000$  & $-0.668{\pm}0.055$ \\  \hline
$100$ & $5000$ &  $-0.723{\pm}0.021$\\ \hline
$200$ & $5000$ & $-0.737{\pm}0.013$ \\ \hline
$400$ & $7900$ & $-0.746{\pm}0.008$ \\ \hline
$600$ & $1000$ & $-0.749{\pm}0.006$ \\ \hline
$800$ & $800$ & $-0.752{\pm}0.005$ \\ \hline
$1000$ & $400$ &  $-0.753{\pm}0.004$\\ \hline
$1600$ & $700$ & $-0.755{\pm}0.003$ \\ \hline
$2000$ & $50$ & $-0.757{\pm}0.002$ \\ \hline
$3000$ & $90$ & $-0.758{\pm}0.002$  \\ \hline
$4000$ & $70$ & $-0.759{\pm}0.001$ \\ \hline
$5000$ & $50$ & $-0.759{\pm}0.001$ \\ \hline
$6000$ & $30$  & $-0.760{\pm}0.001$ \\ \hline
$7000$ & $30$  & $-0.760{\pm}0.001$ \\ \hline
$8000$ & $20$  & $-0.760{\pm}0.001$ \\ \hline
$10000$ & $15$ & $-0.7607{\pm}0.0009$ \\ \hline
    
    \end{tabular}
    \label{tab:table}
\end{table}
\begin{figure*}[tbh]
\includegraphics[width=16cm]{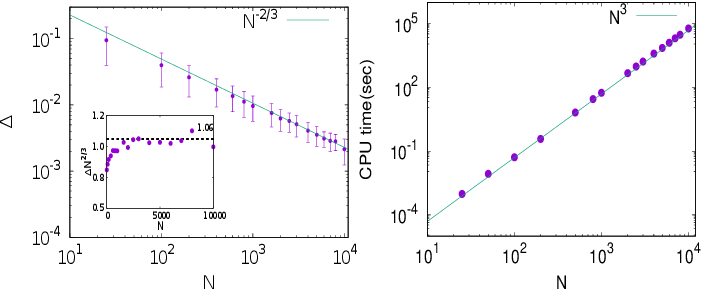}
\caption{The left hand side figure shows the system size dependence of the excess energy from the ground state $\Delta=E^0_N-E^0$ is shown. The power-law variation with exponent $-2/3$ is shown for comparison. The inset shows the pre-factor $a=\Delta N^{2/3}$, which becomes nearly independent of system sizes for larger systems. The right hand side shows the CPU time (time taken to simulate one configuration) goes as $N^3$. }
\label{cc}
\end{figure*}
We compute, at each $t$, the cooperative
energy $E_{N}^{[c]}(t)= - \sum_{<ij>} J_{ij} m^{[c]}_{i}(t) m^{[c]}_{j}(t)$ and the configurational average of $E_{N}^{[c]}(t)$ is denoted by $E_{N}(t) \equiv \left[E_{N}^{[c]}(t)\right]_{av}$. 
This value, at the end of the annealing process ($t=\tau$) would depend, in general, on the annealing protocol. In particular, here it is dependent on the annealing time $\tau$. For example, for a very low value of $\tau$, the local magnetizations would not have sufficient time to adjust to the ground state. However, as $\tau$ is increased, for a particular system size $N$, the energy value will saturate to $E^{0}_{N}$ at $t = \tau_{N}$ (see Fig. \ref{tau_comb}).
For any choice of $\tau>\tau_N$, this saturation value $E^0_N$ does not depend on $\tau$ any more (see inset of Fig. \ref{tau_comb}). It is important to note that the value of $\tau_N$ scales linearly with $N$: $\tau_N \sim N$, as can be seen from the data collapse of the inset in Fig. \ref{tau_comb}. Therefore, for a choice of $\tau$ that varies linearly with $N$ (we have taken it to be $5N$), the classical annealing of the SK model gives the best possible estimate of the ground state energy under the Suzuki-Kubo framework. 

We then proceed to simulate, with a choice of $\tau=5N$, the low energy state $E^{0}_N$ for different system sizes varying from $N=25$ to $N=10000$. The number of configurations used, the final energy values reached after the annealing and the r.m.s. fluctuations in those energy values for different configurations are mentioned in Table \ref{tab:table}. 
As is already indicated in the data collapse in the inset of Fig. \ref{tau_comb}, we obtain a finite size scaling of the low energy states, which shows a scaling form $E^{0}_N-E^0 \sim N^{-2/3}$ (see Fig. \ref{en}) where $E^0$ denotes the actual ground state in the $N \rightarrow \infty$ limit. The fluctuation in $E^0_N$, $\sigma_N\equiv \left[\sqrt{\left <{E^{0}_N}^2\right>-\left<E^{0}_N\right>^2}\right]_{av}$, scales as $\sigma_N\sim N^{-\theta}$ (see inset of Fig. \ref{en}), with $\theta=5/6$ as predicted analytically in \cite{parisi3,parisi4} and numerically seen in \cite{stefan2}, and we also show the fit with $\theta=3/4$, which earlier was the numerical observation.  


Note that the configuration is not necessarily the true ground state
(of any system size $N$), but a low energy configuration whose energy
extrapolates to the actual ground state energy in the large system
size limit. Furthermore, the statistics of these low energy
configurations appear to be the same of the ground state with the
same finite size scaling exponent value (see also \cite{stefan}). The low energy values obtained here (see Table \ref{tab:table}) can be compared with that reported in \cite{stefan2}, where the average energy value is lower. However, due to the lesser algorithmic complexity here, a higher system size could be simulated (see also \cite{selke}).  To quantify this, we have checked, for relatively small system sizes, what fraction of the configurations reach the true ground state using this algorithm. We do this by evaluating the true ground state, by brute force, for small systems and compare that with the low energy states obtained from our algorithm. In Fig. \ref{gnd}, we note that the fraction of cases the true ground state is reached decreases with system size. We also note the probability distribution of the differences of the low energy obtained through our algorithm and the true ground state energy obtained through brute force evaluation. For larger systems, we simulated the system for a given configuration of $J_{ij}$ but with different initial conditions. The system does not reach the exact same state but has a small deviations in the final (after the annealing process) energy values. This also indicates that the configurations reached by our algorithm
are not the true ground state, but their energy is very close to the
ground state energy. Solutions to TAP equations in similar other contexts have indicated the same \cite{cavanga,moore}. 

Given that the system is fully connected, in this algorithm, a single time step requires $N^2$ spin scans. Then, as the annealing time scales linearly with $N$ in order to get the saturation (low) energy value mentioned above, the algorithmic time cost scales as $N^3$ (see Fig. \ref{cc} for the decay of the excess energy  with system size and the CPU time needed to achieve that as a function of the system size).

\section{Discussions and conclusion}
We have reported the results of simulating SK spin glass using mean field Suzuki-Kubo dynamics (Eqs. (\ref{mag}) and (\ref{heff}); the code is available in Ref. \cite{codes}). Unlike the usual convention, where the spins are treated as discrete variable $\pm 1$, the Suzuki-Kubo dynamics make each spin a continuous variable as long as the dynamics continues. Clearly, at the end of the dynamics, where $T=0$, the spins return to their discrete configurations. However, in the initial phase of the dynamics, the spin values are lowered (as evident from the lowering of the spin glass order parameter in Fig. \ref{combine}(b)). This results in a smoothening of the corrugated energy landscape faced by the system, which then is able to quickly orient (within an average annealing time $\tau_N\sim N$; see Fig. \ref{tau_comb}) in a low energy configuration. Subsequently the local magnetization values are increased, so is the spin glass order parameter and the cooperative energy is lowered (as the system had already reached the low energy state when the barriers were small). This mechanism, therefore, gives an overall algorithmic cost of only $N^3$.

Our algorithm, based on the Suzuki-Kubo dynamics \cite{suzuki}(Eq. \ref{mag}), having non-zero $dm_i/dt$ (or practically without any freezing or localization of the individual spins until the end of the annealing process) for continuous values of $m_i$ (and $h_i$) helps much easier approach compared to much more elaborate and involved studied (see e.g., \cite{selke}).
It is important that even in the process of making the individual spin values continuous during the dynamics, the well established finite size scaling of the ground state energy and its fluctuations remain intact (see Fig. \ref{en}). As a result, the low energy states extrapolated for $N\to \infty$ (using results in the range $N=25$ to $N=10^4$) comes to $E^0  = -0.7629\pm 0.0002$, which was earlier obtained (for smaller system sizes) at a higher cost of $N^4$ \cite{boe_epjb2005} (although for a better estimate of the ground state), or the proper finite size scaling ($|E^0_N - E^0|  \sim N^{-2/3}$
and  $\sigma_N \sim N^{-5/6}$) could not be obtained even for much larger system sizes ($N=40000$) \cite{erbal}.

In conclusion, the Suzuki-Kubo dynamics for annealing of SK spin
glass allows to find, in an affordable $O(N^3)$ time, configurations
whose average energy, albeit not being the ground state energy,
extrapolate to it for large $N$.

\section*{Acknowledgements} We are thankful to Muktish Acharyya for useful discussions.
BKC is grateful to the Indian National Science Academy for their
Senior Scientist Research Grant. The simulations were performed using HPCC Chandrama in SRM University-AP. We thank one of the anonymous reviewers for valuable and detailed comments on the earlier version of this manuscript.

\end{document}